\documentclass[proof]{pasj01}

\begin{document}
\SetRunningHead{K. Yasui et al.}{Stellar Density in Galactic Center}

\title{Number Density Distribution of Near-Infrared Sources on 
a
Sub-Degree Scale in the Galactic Center: 
Comparison with the Fe \emissiontype{XXV} K$\alpha$ Line at 6.7 keV}



%
\author{%
   Kazuki \textsc{Yasui}\altaffilmark{1}, %
   Shogo \textsc{Nishiyama}\altaffilmark{2},
   Tatsuhito \textsc{Yoshikawa}\altaffilmark{1},
   Schun \textsc{Nagatomo}\altaffilmark{1},
   Hideki \textsc{Uchiyama}\altaffilmark{3},
   Takeshi Go \textsc{Tsuru}\altaffilmark{4},
   Katsuji \textsc{Koyama}\altaffilmark{4, 5},
   Motohide  \textsc{Tamura}\altaffilmark{6},
   Jungmi \textsc{Kwon}\altaffilmark{6},
   Koji \textsc{Sugitani}\altaffilmark{7},
   Rainer \textsc{Sch\"{o}del}\altaffilmark{8},
   and
   Tetsuya \textsc{Nagata}\altaffilmark{1}}
 \altaffiltext{1}{Department of Astronomy, 
Graduate School of Science, 
Kyoto University, Kyoto 606-8502}
 \email{yasui@kusastro.kyoto-u.ac.jp, nagata@kusastro.kyoto-u.ac.jp}
 \altaffiltext{2}{Miyagi University of Education, Aoba-ku, Sendai 980-0845
}
 \altaffiltext{3}{Science Education, Faculty of Education, Shizuoka University, 
Shizuoka 422-8529}
 \altaffiltext{4}{Department of Physics, Graduate School of Science, Kyoto University, Kyoto 606-8502}
 \altaffiltext{5}{Department of Earth and Space Science, Graduate School of Science, Osaka University, 
Osaka 560-0043}
 \altaffiltext{6}{Department of Astronomy, Graduate School of Science, The University of Tokyo, Tokyo 113-0033}
 \altaffiltext{7}{Graduate School of Natural Sciences, Nagoya City University, Nagoya 467-8501}
 \altaffiltext{8}{Instituto de Astrofisica de Andalucia (CSIC), Glorieta de la Astronomia s/n, 18008, Granada, Spain }


\maketitle

\begin{abstract}
%
The stellar distribution derived from an $H$ and $K_{\mathrm S}$-band survey  
of the central region of our Galaxy 
is compared with the Fe \emissiontype{XXV} K$\alpha$ (6.7 keV) line intensity 
observed with the Suzaku satellite. 
The survey is for the Galactic coordinates 
$|l| \lesssim 3\arcdeg.0$ and $|b| \lesssim 1\arcdeg.0$
(equivalent to 0.8\,kpc $\times$ 0.3\,kpc for $R_0 = 8$\,kpc), 
and 
the number-density distribution $N(K_{\mathrm S,0}; l, b)$ of stars 
is derived 
using 
the extinction-corrected magnitude $K_{\mathrm S,0}=10.5$.  
This is deep enough to probe the old 
red giant 
population 
and in turn to estimate the ($l$, $b$) distribution of faint X-ray point sources 
such as coronally active binaries and cataclysmic variables.  
In the Galactic plane ($b=0\arcdeg$), 
$N(10.5; l, b)$ increases to the Galactic center as $|l|^{-0.30 \pm 0.03}$ 
in the range of $-0\arcdeg.1 \geq l \geq -0\arcdeg.7$, 
but this increase is significantly slower than 
the increase ($|l|^{-0.44 \pm 0.02}$ ) of the 
Fe \emissiontype{XXV} K$\alpha$ line intensity.  
If normalized with the ratios in the outer region  
$1\arcdeg.5 \leq |l| \leq 2\arcdeg.8$, 
where faint X-ray point sources are argued to 
dominate the diffuse Galactic X-ray ridge emission, 
the excess of the Fe \emissiontype{XXV} K$\alpha$ line intensity over the stellar number density is 
at least a factor of two
at $|l| = 0\arcdeg.1$.  
This indicates that 
a significant part of the Galactic center diffuse emission arises  
from a truly diffuse optically-thin thermal plasma, 
and not from an unresolved collection of faint X-ray point sources 
related to the old stellar population.
\end{abstract}

\section{Introduction}
The Galactic center diffuse X-rays (GCDX) 
were found as a flux peak having 
strong emission lines around 6.7 keV 
with a spatial width of $\sim 1\arcdeg$ 
\citep{koy89} 
in the Galactic ridge X-ray emission (GRXE).  
The emission was resolved into three lines at 
6.4, 6.7, and 7.0 keV 
arising from neutral, He-like, and H-like iron \citep{koy96}, 
respectively.   
The 6.7 keV K$\alpha$ line from He-like iron (Fe \emissiontype{XXV}) is regarded 
as a pure thin thermal plasma component, and 
its spatial distribution has been studied by 
\citet{uch11}.  
The spatial profile 
in the GRXE region ($|l| >$ 1.$\arcdeg$5)
was 
well fitted by a stellar mass distribution model, 
but the profile 
in the GCDX region (0.$\arcdeg$2 $< |l| <$ 1.$\arcdeg$5) showed fourfold excess.  
\citet{uch11} therefore concluded that 
the Fe \emissiontype{XXV} K$\alpha$ line in the GCDX 
could not 
be explained in the same way as the GRXE.  
The GRXE seems to be dominated by faint X-ray point sources such as
coronally active binaries (ABs) and cataclysmic variables (CVs) 
\citep{rev06b,rev09,yua12}.  
These sources belong to the old ($\gtrsim$Gyr) stellar population, 
and their X-ray luminosity function normalized 
to the stellar mass is not expected to vary significantly across
the Galaxy 
\citep{saz06}.

\citet{uch11} 
compared 
the spatial profile of the Fe \emissiontype{XXV} K$\alpha$ line 
with the stellar mass distribution of the nuclear bulge (NB)
of the Galaxy, 
following \citet{mun06}; 
they both employed the 
elaborate stellar mass distribution model of the NB 
(distance from the Galactic center 
$R \lesssim 300 \mathrm{pc}$ 
or $|l| \lesssim 2\arcdeg$) 
by 
\citet{phi99}, \citet{mez99}, and \citet{lau02},
together with additional 
information about the more extended part of 
the Galactic bulge (GB; 
$R \lesssim 3 \mathrm{kpc}$ 
or $|l| \lesssim 20\arcdeg$) 
and disk 
\citep{ken91}.  
The NB model is an empirical 3-D axisymmetric model 
consisting of a spherical nuclear stellar cluster and a nuclear stellar disk.  
It is mainly based on the COBE 4.9\,$\mu$m surface brightness map with an angular resolution of 0.$\arcdeg$7.  
Since the NB is unresolved in Galactic latitude, 
the scale height of warm dust derived from IRAS data was substituted for the 
stellar disk.  
\citet{uch11} and \citet{mun06}
in turn integrated the NB (and GB) model along the line of sight 
toward the Galactic center 
while tuning the relative weight of each component, and 
compared these model surface densities with the observed surface density of 
the Fe \emissiontype{XXV} K$\alpha$ line.  
\citet{mun06}, however, 
estimated the overall uncertainty in building this model to be as large as 
50\%.  

Here we take another approach to this problem: directly comparing  
the observed surface density of 
the Fe \emissiontype{XXV} K$\alpha$ line with 
the observed (and extinction-corrected) stellar number density $N (l, b)$ 
in the $K_{\mathrm S}$ band, 
without making use of of any models for the NB or GB.  
Since the NB and GB are highly concentrated\footnote{
A
simple calculation based on 
the \citet{wai92} model shows that only $\sim$7 percent stars 
along
the line of sight $(l=2\arcdeg, b=0)$ 
are located 
at a distance of 
7.5\,kpc or less from the Sun. } 
toward the Galactic center, 
the line of sight depth can be neglected, to 0th order.  
Therefore, all the stars, as well as the Fe \emissiontype{XXV} K$\alpha$ emission, can be considered to be located 
at a distance of 8\,kpc.  
Then the absolute magnitude $M_{K_{\mathrm S}}$ and extinction-corrected apparent magnitude $K_{\mathrm S,0}$ 
of stars 
are related simply by 
$M_{K_{\mathrm S}} +14.52 = K_{\mathrm S,0}$,
where 14.52 is the distance modulus.  
We aim to probe the ($l$, $b$) distribution of 
the 
old population 
to which the faint X-ray point sources belong, by 
counting stars more luminous than $M_{K_{\mathrm S}}$ 
or, equivalently, 
brighter than the apparent magnitude 
$K_{\mathrm S,0}$.  
As a first assumption, 
if we set the limiting magnitude 
to be 
reasonably faint, 
we will be able to detect a representative old 
red giant 
population 
and 
the derived ($l$, $b$) distribution 
$N(K_{\mathrm S,0}; l, b)$ 
should be similar to that of faint X-ray point sources.  
The earliest and faintest M giants have $M_{K_{\mathrm S}}=-4.14$ \citep{wai92}.  
We thus require 
a
corresponding limiting magnitude of 
$K_{\mathrm S,0}$ $\sim$10.4.  
If 
we cannot set the limiting magnitude faint enough, 
we will end up detecting only supergiants and other younger stars, 
whose spatial distribution can be very different.
%
In this {\it Paper}, 
employing large-scale near-infrared survey data with faint limiting magnitudes,  
we derive the number density map of stars to the extinction-corrected magnitude $K_{\mathrm S,0}$ 
deep enough
to detect the old red giant population, 
and estimate the ($l$, $b$) distribution of faint X-ray point sources.  
%
%
%

\section{Observation and Data Analysis}
%
%
%
The central region of our Galaxy $|l| \lesssim$ $3\arcdeg.0$ and $|b| \lesssim 1\arcdeg.0$
(equivalent to 0.8\,kpc $\times$ 0.3\,kpc for $R_0 = 8$\,kpc) 
was observed \citep{nis06}  
using the near-infrared 
camera SIRIUS (Simultaneous Infrared Imager for Unbiased Survey; 
\cite
{nag99, nag03})
on the 1.4m IRSF telescope. 
The SIRIUS 
camera provides $J$ (1.25\,$\mu$m), $H$ (1.63\,$\mu$m), and $K_{\mathrm S}$
(2.14\,$\mu$m) images simultaneously, with a field of view of 7$'$.7 $\times$
7$'$.7 and a pixel scale of 0$''$.45.
The typical seeing was 1$''$.2 FWHM, 
and 
the averages of the 10-$\sigma$ limiting
magnitudes were $J$ = 17.1, $H$ = 16.6, and $K_{\mathrm S}$ = 15.6.  
We will employ only the $H$ and $K_{\mathrm S}$ data 
because of heavy extinction toward the Galactic center.  

\begin{figure}
  \begin{center}
    \FigureFile(80mm,100mm){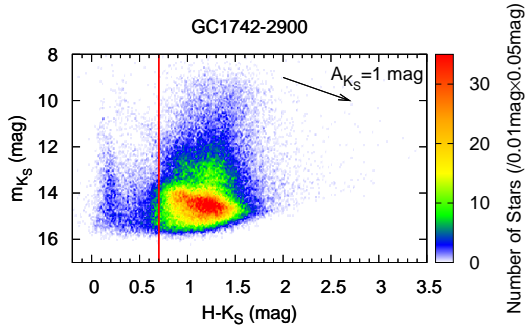}
  \end{center}
\caption{The color magnitude diagram ($H-K_{\mathrm S}$ vs $K_{\mathrm S}$) of the 20\,$'$ $\times$ 20\,$'$ region 
         centered at $(l,b) = (-0\arcdeg.41, 0\arcdeg.53)$. 
         Stars to the right of red line are selected as the stars near the Galactic center.  
These stars are dereddened oppositely to the arrow based on \citet{nis06}, 
to the intrinsic color of red giants $(H-K_{\mathrm S})_0=0.2$.}
\label{fig:cmd}
\end{figure}

We choose stars near the Galactic center, 
and derive their extinction-free magnitudes 
from $H-K_{\mathrm S}$ vs $K_{\mathrm S}$ color magnitude diagrams (CMDs).  
Three by three fields of view were treated as one region of 
20\,$'$ $\times$ 20\,$'$.  
Figure \ref{fig:cmd} is the CMD of a region
centered at $(l,b) = (-0\arcdeg.41, 0\arcdeg.53)$, shown as an example.  
Estimates based on a simple model \citep{wai92} 
show 
that 
the majority of detected stars are near the Galactic center 
and reddened to the color 
corresponding to $H-K_{\mathrm S} \gtrsim 0.5$.  
Therefore, the stars near the Galactic center are selected by eye in the CMDs, 
and 
then their $K_{\mathrm S}$ magnitudes were corrected for extinction.  
The threshold of $H-K_{\mathrm S}$ was 
0.7 for the region shown in Figure 1, and was 
in the range of $0.35-1.10$ for each region of the whole observation.  
The average intrinsic color $(H-K_{\mathrm S})_0$ of the detected stars 
on the basis of 
calculations, 
considering the limiting magnitudes 
and using the 
\citet{wai92} model,  
is 0.203;  
its standard deviation 
is quite small because the intrinsic color $(H-K_{\mathrm S})_0$ of stars is 
0.0 
for the 
bluest (rare OB stars) and 0.4 
for the 
reddest (rare late M giants).  
Therefore, we regard the intrinsic color $(H-K_{\mathrm S})_0$ of all stars as 0.203, and 
corrected their $K_{\mathrm S}$ magnitudes according to the reddening law by
\citet{nis06} 
$A(K_{\mathrm S}) = 1.44 \times [
(H-K_{\mathrm S}) - 
(H-K_{\mathrm S})_0
]
$
;   
see 
the arrow in 
Figure
\ref{fig:cmd}.  
Thus we obtain the line-of-sight extinction $A(K_{\mathrm S})$ and extinction-corrected magnitude $K_{\mathrm S,0}$ 
of each star.

There has been considerable work carried out on the near-IR extinction in the Galactic bulge. 
Some of 
the work 
indicates
that the wavelength dependence of near-IR extinction varies 
sufficiently 
from one line of sight to another that 
the use of a single extinction law can lead to 
an incorrect
stellar distribution.  
\citet{nat13} derived $I$-band extinction laws as a function of 
two 
colors of ($J-K_S$) and ($V-I$), 
and concluded that the conversion 
from reddening to extinction cannot be done accurately with a
single 
law.  
\citet{che13} found large variations in extinction over angular scales 
as small as $15' \times 15'$ .
Furthermore, they found `grayer' extinction coefficient $A_{\lambda} / A_{K_S}$
toward the inner Galaxy,
where $\lambda$ is the Spitzer IRAC bands of 3.6, 4.5, 5.8, and 8.0$\mu$m.    
We have checked the 
impact
of such 
a
variation, 
assuming that the $H$-band extinction 
coefficient changes as much as the 3.6$\mu$m coefficient.  
While the {\it variation} in $A_{3.6} / A_{K_S}$ over the central several degrees 
of longitude seems one percent at most (see their Fig.24(a)), 
this variation affects the dereddening procedure and leads to 
a several percent change in the number count.
Thus we should be careful if we discuss more subtle differences.

We set the limiting 
magnitude $K_{\mathrm S,0}=10.5$, and 
plotted 
the stellar number density profile $N(10.5; l, b)$.  
This 
corresponds to the absolute magnitude of $M_{K_{\mathrm S}} = -4.0$, 
which includes all M-type giants 
at the Galactic center.  
They are numerous enough 
(nearly two orders of magnitude more than young supergiants in the 
\citet{wai92} model) 
to probe the old stellar population, and we therefore assume that 
$N(10.5; l, b)$ traces adequately the spatial distribution of faint X-ray point sources.   
In fact, 
\citet{nis13} use 
a
synthetic CMD computation \citep{apa04} 
to estimate the fraction of the stars brighter than  $K_{\mathrm S,0}=10.5$ 
for
several different models of star forming history; 
they found that these bright stars trace 
the spatial distribution of the faint old stars 
very well, even in a conservative case.  

\begin{figure}
  \begin{center}
  \FigureFile(8cm,8cm){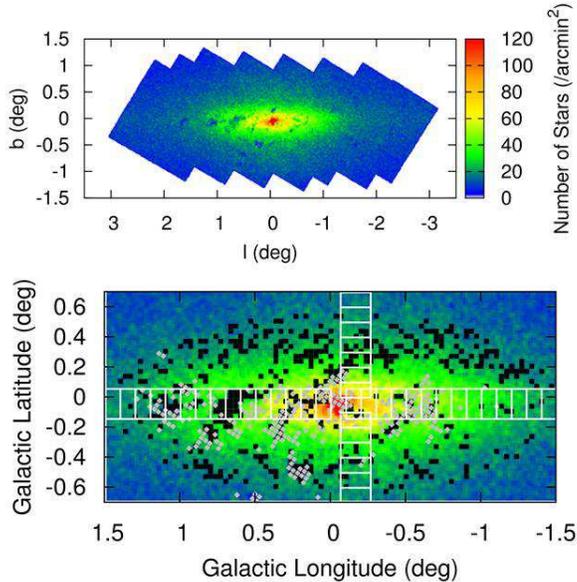}
  \end{center}
\caption{(a)    
The number density $N(10.5; l_*, b_*)$ (${\rm arcmin^{-2}}$) of stars 
whose extinction-corrected $K_{\mathrm S}$ are brighter than 10.5.  
(b)
Close-up of the central region. 
The same fields of view  
as \citet
{uch11}, where 
the Fe \emissiontype{XXV} K$\alpha$ line intensity and 
$N(10.5; l_*, b_*)$ 
are plotted in 
Figure 
\ref{fig:lprof}, are shown by {\it white}-outlined rectangles.   
Line of sight extinction derived from Spitzer/IRAC observations
$A(K_{\rm S, Spitzer})$ is greater than 3 in {\it gray} squares, and  
stellar distribution 
evidently decreases 
from the adjacent regions 
in {\it black} squares.
These squares are removed from further analysis.}
\label{fig:map}
\end{figure}

Since our 10-$\sigma$ limiting magnitudes are $ (H,K_{\mathrm S}) = (16.6, 15.6)$ and 
the extinction is $A(K_{\mathrm S}) \sim 2-3.5$\,mag 
even for the Galactic center itself 
\citep[the 40\,$''$ $\times$ 40\,$''$ region around Sgr\,A$^*$]{sch10}, 
we 
can 
assume that 
our star list is nearly complete to the extinction-corrected magnitude $K_{\mathrm S,0}=10.5$
over almost all the regions near the Galactic center.  

First, we examine this assumption.  
We construct an extinction-free map of the stellar number density $N(10.5; l, b)$ 
with a 1$'$ resolution in 
Figure 
\ref{fig:map}(a).  
Using these bright stars of $K_{\mathrm S,0}<10.5$, 
we can estimate the average extinction 
towards these stars at a resolution of 
1$'$ as well.  
%
We can 
check if the extinction derived 
in 
this way is consistent with 
the extinction
$A(K_{\rm S, Spitzer})$ derived from Spitzer/IRAC observation between 3.6 and 8.0\,$\mu$m 
\citep{sch09}.  
Since the Spitzer/IRAC extinction values were derived 
from longer wavelength observations, 
they tend to represent the line-of-sight extinction to the Galactic center better 
in more reddened regions.  
The comparison of these extinction values in 
Figure 
\ref{fig:aks} 
indicates that our estimates agree very well with those from Spitzer/IRAC 
up to the extinction of $A(K_{\rm S, Spitzer}) \sim 3$.  
In some of the more reddened regions, however, our extinction is smaller than Spitzer/IRAC. 
Therefore, we mark the map pixels whose $A(K_{\rm S, Spitzer}) > 3$ by gray squares in 
Figure 
\ref{fig:map}(b).  
Figure \ref{fig:map}(b) has another type of squares: 
the black squares are the 2$'$ $\times$ 2$'$ regions (four pixels combined) where 
our stellar distribution 
evidently decreases 
from the adjacent regions. 
To select these pixels, we have calculated the running mean of the stellar density 
in 15 regions of 2$'$ $\times$ 2$'$ in $l$ and $b$.  
These lines of sight might suffer so great 
an 
extinction that 
even $A(K_{\rm S, Spitzer})$ might have been underestimated.  
These two types of pixels tend to increase toward the central region, and 
this leads to underestimation of the stellar number density in the central region. 
The masked pixels were removed from further analysis, 
which amount to 18 percent of the area of Figure \ref{fig:map}(b).

\begin{figure}
  \begin{center}
  \FigureFile(6cm,12cm){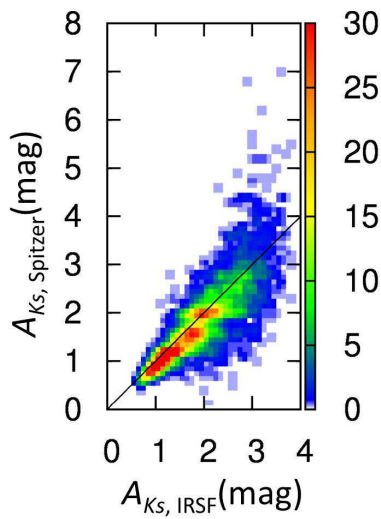}
  \end{center}
  \setcounter{figure}{2}
\caption{The extinction derived from IRSF data (current work) and that derived from Spitzer/IRAC data
\citep{sch09}.  
The regions whose $A(K_{\rm S, Spitzer}) > 3$ deviate from the straight line, and are shown as gray squares 
in 
Figure 
\ref{fig:map}(b).}
\label{fig:aks}
\end{figure}

Second, we estimate the effect of confusion on 
finding stars.  
The stellar density increases by a factor of $\sim\,5$ from the outer part of our data 
to the central $0\arcdeg.1$.  
We computed the completeness of finding stars over the entire field. 
We added artificial sources of $K_{\mathrm S}=12.5$ 
to the SIRIUS images, and examined the recovery rate of these sources (see also
\cite{hat13}).   
It is generally quite high $\sim 97\%$ in the outer part, 
but drops to $\sim 92\%$ in the center.  
Therefore, the stellar number density 
is slightly underestimated in the central part, and 
this distorts the profile slightly.  
We will correct for these completeness changes in the analysis in the next section.  

To check further whether  
the stellar number density 
$N(10.5; l, b)$ 
is not severely underestimated in 
the central part of our map, 
we 
constructed 
the number density $N(8.0; l, b)$ of stars brighter than 
$K_{\mathrm S,0} = 8.0$
because brighter stars are 
less affected by confusion and extinction problems, 
which generally increases toward the center.  
We test if $N(10.5; l, b)$ and $N(8.0; l, b)$ show similar profiles.  
In the outer part, the two generally agree within the Poisson errors,   
although $N(8.0; l, b)$ is sometimes too small to make any meaningful comparison.  
In the central part, 
we calculate the averages of 
$N(10.5; l, b) / N(8.0; l, b)$ in 
the four 20\,$'$ $\times$ 20\,$'$ regions 
adjacent to the central 20\,$'$ $\times$ 20\,$'$ region.  
The ratio 
is quite constant $\sim7.3$, with 
a region to region variance of 0.6.   
In the eight
7$'$ $\times$ 7$'$ sub-regions in the central 20\,$'$ $\times$ 20\,$'$ 
region\footnote{One region of 20$'$ $\times$ 20$'$ consists of nine 7$'$ $\times$ 7$'$ sub-regions, 
but the very central sub-region is so crowded 
with the stars belonging to the central-parsec cluster, 
we did not derive the stellar density reliably. See \citet{nis13}.}, 
when we multiply $N(8.0; l, b)$ by this ratio, 
the results are greater than $N(10.5; l, b)$ 
by factors $0.97-1.50$.
The mean of these factors is 1.21, 
i.e., 
greater than 1.00; this could be due to 
1) the extinction or confusion mentioned above, or 2) 
actual change in the bright-to-faint star ratio.  
The latter is possible if the central part of NB 
contains younger stars which are often more luminous.  
If we were affected by this effect, 
then we would overestimate the point-source-originated Fe \emissiontype{XXV} K$\alpha$ emission 
from $N(8.0; l, b)$ 
in the central region.  
However, we would certainly be able to determine reliable upper limits.  
Therefore, 
we will use these corrected 
$N(10.5; l, b)$, derived from $N(8.0; l, b)$, in the central 20$'$ region 
in the next section.  

%

\section{Results and Discussion}

The stellar number density distribution is compared with 
that of 
the 
Fe \emissiontype{XXV} K$\alpha$ line obtained 
with the X-ray Imaging
Spectrometer onboard the Suzaku satellite 
reported 
in \citet{uch11}.  
We will use the new coordinates whose origin is at Sgr A$^*$, 
$(l_*, b_*) = (l + 0\arcdeg.056, b + 0\arcdeg.046)$,  
following 
\citet{uch11}.  
The Fe \emissiontype{XXV} K$\alpha$ intensity shows 
several tens of percent excess 
on the positive $l_*$ side (east)
in comparison with the negative side (west) near the center 
$0\arcdeg.1\leq |l_*| \leq 0\arcdeg.3$ 
\citep{koy07, koy09, uch11}.  
This asymmetry was also pointed out in the number of 
X-ray stars by \citet{mun09}.  
They suggest that the excess is due to 
young massive stars associated with the high activity of star formation 
in the positive $l_*$ part.  
In contrast, our 
measurement of the 
near-infrared stellar number density 
shows no significant asymmetry, as shown in Figure \ref{fig:lprof}(b).
Here we compare the Fe \emissiontype{XXV} K$\alpha$ intensity and 
the near-infrared stellar number density 
on the negative $l_*$ side where the former increases 
towards the center 
at a slower rate. 

The number density $N(l_*, b_*)$ in 
the same rectangular bins 
as \citet
{uch11}, 
$(\Delta l_*, \Delta b_*) = (0\arcdeg.1, 0\arcdeg.2) $ for the longitude distribution 
is plotted in 
Figure 
\ref{fig:lprof}(a), 
and 
$(\Delta l_*, \Delta b_*) = (0\arcdeg.2, 0\arcdeg.1) $ for the latitude distribution, 
is in 
Figure 
\ref{fig:lprof}(b).  
The center of each rectangle is given as 
$(l_*, b_*) = (0\arcdeg.1 \times n, 0.)$ along the longitude, and   
$(l_*, b_*) = (-0\arcdeg.114, 0\arcdeg.1 \times n)$ along the latitude,    
where $n$ is a non-zero integer.  
The number density in  each rectangular bin corresponds to the average in the unmasked $1' \times 1'$ pixels 
shown in 
Figure 
\ref{fig:map}(b), 
and the averaged location of them is also calculated for that specified rectangle.   
In 
Figure 
\ref{fig:lprof}(a) and (b), we normalize the stellar number density so that the ordinates agree 
in the GRXE region $(1\arcdeg.5 \leq |l_*| \leq 2\arcdeg.8, b_* = 0\arcdeg.0)$, 
where \citet{rev06b} found that the Fe \emissiontype{XXV} K$\alpha$ line intensity can be explained by X-ray point sources.
The Fe \emissiontype{XXV} K$\alpha$ line intensity shows significant excess in comparison with 
the stellar number density in the central region, and 
the excess increases as we approach the center.  
The excess is $\sim 1.5$ at $l_*=-0\arcdeg.7$, and reaches $\sim 2$ at $l_*=-0\arcdeg.1$ 
when normalized in the Galactic ridge.  
The profiles in the Galactic latitude show the same tendency
as shown in 
Figure 
\ref{fig:lprof}(b).

\begin{figure}
  \begin{center}
  \FigureFile(8cm,6cm){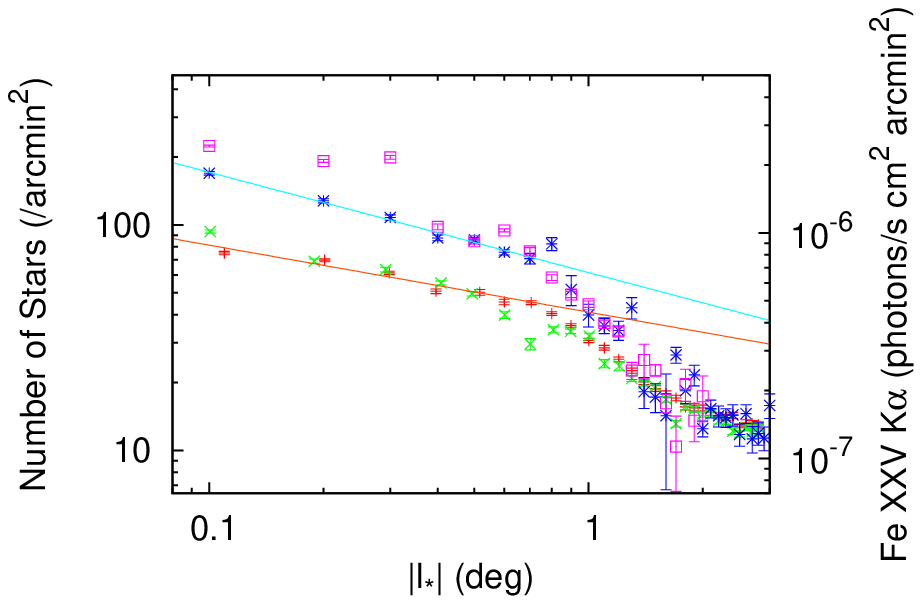}
  \end{center}
  \setcounter{figure}{3}
\caption{(a)
The stellar number density distribution $N(10.5; l_*, b_*) ({\rm arcmin^{-2})}$ 
along the Galactic plane as a function of $l_*$ (distance from Sgr A$_*$), 
{\it red}: negative Galactic longitude and {\it green}: positive.  
The error bars are Poisson errors, and include the uncertainty in the conversion from 
$N(8.0; l_*, b_*)$ for the central two positions.  
Also shown is 
Fe \emissiontype{XXV} K$\alpha$ line intensity distribution $({\rm photons\,\, s^{-1} cm^{-2} arcmin^{-2}})$, 
{\it blue}: negative Galactic longitude and {\it magenta}: positive. 
Note that 
the {\it blue} and {\it magenta} crosses are exactly at $|l_*|=0.1, 0.2, ...$, while 
the {\it red} and {\it green} crosses are slightly off because the averaged position in 
the rectangles in 
Figure 
\ref{fig:map}(b) is not necessarily their centers 
because of the exclusion of gray and black squares
in 
Figure 
\ref{fig:map}(b).   
The ordinates are set so that 
Fe \emissiontype{XXV} K$\alpha$ emission and $N$ agree 
in the region $-1\arcdeg.5 \geq l_* \geq -2\arcdeg.8$. 
Power-law approximation 
in the region $-0\arcdeg.1 \geq l_* \geq -0\arcdeg.7$ 
yields the indices of 
$-0.44\pm 0.02$
and
$-0.30\pm 0.03$.
}
\label{fig:lprof}
  \begin{center}
  \FigureFile(8cm,6cm){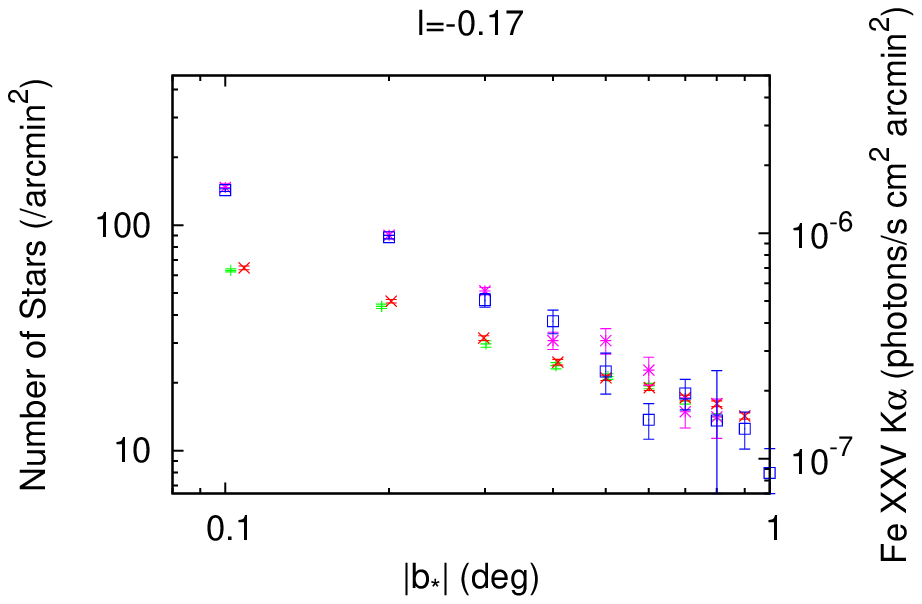}
  \end{center}
\setcounter{figure}{3}
\caption{(b)
Same as (a), along the Galactic longitude of $l_*= -0\arcdeg.114$ 
as a function of $b_*$.  
The crosses are 
{\it magenta}: Fe \emissiontype{XXV} K$\alpha, l* > 0$,
{\it blue}: Fe \emissiontype{XXV} K$\alpha, l* < 0$,
{\it green}: $N(10.5), l_* > 0$,
{\it red}: $N(10.5), l_* < 0$.
}
\label{fig:bprof}
\end{figure}

$N(10.5; l_*, b_*)$ increases to the Galactic center as $|l_*|^{-0.30 \pm 0.03}$ at $b_*=0\arcdeg$ 
in the range of $-0\arcdeg.1 \geq l_* \geq -0\arcdeg.7$, 
and 
the increase of the Fe \emissiontype{XXV} K$\alpha$ line intensity in \citet{uch11} does as $|l_*|^{-0.44 \pm 0.02}$.  
The difference in these indices are independent of the region we choose for the normalization.  
Therefore, the Fe \emissiontype{XXV} K$\alpha$ line is likely to originate from emitters with 
different spatial distribution from the old population.  

Our excess estimate is $\sim2$ at  $l_*=-0\arcdeg.1$, 
but 
\citet{uch11} concluded that
the emissivity in the NB must be higher by a factor of 
$3.8\pm0.3$ 
than that in the GRXE 
when fitted with stellar mass distribution models.  
If we look at their Fig. 4
(similarly the lower-right panel of Fig. 7 of
\cite{hea13}),
the stellar mass distribution curve based mainly on 
\citet{lau02} 
seems to be very flat compared with the Fe \emissiontype{XXV} K$\alpha$ distribution, 
especially in the range of $0\arcdeg.1 \lesssim |l_*| \lesssim 0\arcdeg.7$. 
The stellar mass model assumes a nuclear stellar disk whose radius is about 150\,pc, 
and the projected surface density is rather flat.  
In contrast, our stellar number density increases gradually toward the center.   
Its increase from $l = 0\arcdeg.7$ to $l = 0\arcdeg.1$ is 
about 1.4 times 
that of 
the flatter mass model by \citet{lau02}.  
This difference in the stellar number density profile likely explains part of the large emissivity difference 
derived from our analysis and that of \citet{uch11}.  
Also, we have normalized the infrared/X-ray ratio 
in the range of $1\arcdeg.5 \leq |l_*| \leq 2\arcdeg.8$ 
with a slightly greater value than 
\citet{uch11} 
because we have not included 
lower 
X-ray 
intensities 
around $l = 8\arcdeg$ in 
Fig. 4 of Uchiyama et al. 
(Data by \cite{yam09}).  
This leads to slightly smaller value in the current estimate than 3.8 in 
\citet{uch11}.  

Another remarkable feature of 
Figure 
\ref{fig:lprof}(a) is the apparent change in these slopes 
around $l_*=0\arcdeg.8$ both in $N(10.5; l_*, b_*)$ and the Fe \emissiontype{XXV} K$\alpha$ line intensity.  
Within 
this Galactic longitude, the nuclear disk component is dominant 
in the stellar mass distribution model of the NB 
\citep{lau02}.  
This indicates that 
the Fe \emissiontype{XXV} K$\alpha$ line excess is related to the nuclear disk, and more concentrated to the center 
than its usual stellar component.  

Finally, we discuss the possible systematic errors of the measured stellar number density.  
Although we corrected for the change in star-finding completeness and 
adopted the larger $N(10.5; l_*, b_*)$ derived from brighter stars of $K_{\mathrm S,0}<8.0$, 
the data point at $l_*=-0\arcdeg.1$ shows a slight downward shift from the straight line 
in 
Figure 
\ref{fig:lprof}(a).  
To be conservative, we tried 
fitting without this data point, 
and obtained 
$-0.36 \pm 0.03$ 
as the index of  
stellar number density distribution.  
When the point at $l_*=-0\arcdeg.1$ is excluded, 
the index of the Fe \emissiontype{XXV} K$\alpha$ distribution is
$-0.47 \pm 0.03$.  
Thus even with these exclusions, 
the slopes seem to be different.  

Thus, the diffuse X-ray emission characterized by the Fe \emissiontype{XXV} K$\alpha$ lines 
in the range of $|l_*| \lesssim$ 3.$\arcdeg$0 and $|b_*| \lesssim$ 1.$\arcdeg$0
most likely arises 
from 1) a point-source component which is detected as old 
red giants 
in our infrared survey 
and 
extends to the outer part of the GB, 
and 2) a more concentrated component which becomes prominent inside $|l_*| \leq$ 0.$\arcdeg$8.    
The second component might be related to the NB of the Galaxy, 
where active star formation takes place 
(see, e.g., 
\cite{mat11}
).  
Supernova explosions of newly born stars might be able to provide hot plasma, 
but the confinement of such plasma is another difficult problem, which was 
discussed in \citet{nis13}.

\section{Summary}

Direct comparison of the observed surface density of the Fe \emissiontype{XXV} K$\alpha$ line with 
the extinction-corrected stellar number density observed in the $K_{\mathrm S}$ band 
yields different power-law indices 
$|l_*|^{-0.44 \pm 0.02}$ and 
$|l_*|^{-0.30 \pm 0.03}$, respectively, 
in the range of $-0\arcdeg.1 \geq l_* \geq -0\arcdeg.7$.
The Fe \emissiontype{XXV} K$\alpha$ line intensity shows significant excess in comparison with 
the stellar number density in the central region: 
$\sim 1.5$ at $l_*=-0\arcdeg.7$, and reaches $\sim 2$ at $l_*=-0\arcdeg.1$
when normalized in the Galactic ridge.  
Therefore, 
we conclude that 
the diffuse X-ray emission in the Galactic Center region cannot be explained 
solely by a population of unresolved point-sources that is related to the old stellar population.


\bigskip
We are grateful to the SAAO staff for their outstanding support of the operations of IRSF.  
We would also like to thank J. H. Hough for very helpful comments. 
This work was
partly supported by the Grants-in-Aid
Scientific Research (C) 21540240, and the Global
COE Program ``The Next Generation of Physics, Spun from
Universality and Emergence'' from the Ministry of Education,
Culture, Sports, Science and Technology (MEXT) of Japan.
RS acknowledges support 
by the Ram\'on y Cajal programme and 
by grants AYA2010-17631 and AYA2009-13036 of 
the Spanish Ministry of Economy and Competitiveness and 
by grant P08-TIC-4075 of the Junta de Andaluc\'ia. 
SN acknowledges support by 
the Grant-in-Aid for Specially Promoted Research 22000005, and
Grant-in-Aid for Young Scientists (A) 25707012.  
We warmly thank 
the anonymous referee for helpful comments 
which improved the presentation of this paper.



\end{document}